\RequirePackage{lineno}
\documentclass[twocolumn,showpacs,superscriptaddress,amsmath,amssymb]{revtex4-1}
\usepackage[colorlinks,linkcolor=blue,anchorcolor=blue,citecolor=blue,urlcolor=blue]{hyperref}
\usepackage{graphicx}
\usepackage{epsfig}
\usepackage{dcolumn}
\usepackage{bm}
\usepackage{overpic}
\usepackage{color}
\usepackage{multirow}
\usepackage{subfigure}
\usepackage{ulem}
\newcommand{\ks}{K_{S}^{0}}

\newcommand{\ee}{e^{+}e^{-}}

\newcommand{\pipi}{\pi^{+}\pi^{-}}
\newcommand{\gaga}{\gamma\gamma}

\newcommand{\sqs}{\sqrt{s}}

\newcommand{\pio}{\pi^{0}}

\newcommand{\Nhobs}{N_{h}^{\textmd{obs}}}

\newcommand{\Npioobs}{N_{\pi^{0}}^{\textmd{obs}}}
\newcommand{\Nksobs}{N_{\ks}^{\textmd{obs}}}

\newcommand{\Nbarhobs}{\bar{N}_{h}^{\textmd{obs}}}

\newcommand{\Nbarhtru}{\bar{N}_{h}^{\textmd{tru}}}

\newcommand{\Nhadtot}{N_{\textmd{had}}^{\textmd{tot}}}
\newcommand{\Nhadobs}{N_{\textmd{had}}^{\textmd{obs}}}
\newcommand{\Nbarhadobs}{\bar{N}_{\textmd{had}}^{\textmd{obs}}}

\newcommand{\Nbarhadtru}{\bar{N}_{\textmd{had}}^{\textmd{tru}}}

\newcommand{\Nbkg}{N_{\textmd{bkg}}}

\newcommand{\gev}{\mathrm{GeV}}
\newcommand{\mev}{\mathrm{MeV}}

\newcommand{\gevc}{\mathrm{GeV}/c}

\let\oldequation\equation
\let\oldendequation\endequation

\renewenvironment{equation}
{\linenomathNonumbers\oldequation}
{\oldendequation\endlinenomath}

\begin{document}
\title{\boldmath Measurements of Normalized Differential Cross Sections of Inclusive $\pio$ and $\ks$ Production in $\ee$ Annihilation at Energies from 2.2324 to 3.6710 $\gev$}

%% Saved at => 2022-07-21
\author{
\begin{small}
\begin{center}
M.~Ablikim$^{1}$, M.~N.~Achasov$^{12,b}$, P.~Adlarson$^{72}$, M.~Albrecht$^{4}$, R.~Aliberti$^{33}$, A.~Amoroso$^{71A,71C}$, M.~R.~An$^{37}$, Q.~An$^{68,55}$, Y.~Bai$^{54}$, O.~Bakina$^{34}$
% ...
\\
\vspace{0.2cm}
(BESIII Collaboration)\\
\vspace{0.2cm} {\it
$^{1}$ Institute of High Energy Physics, Beijing 100049, People's Republic of China\\
% ...
$^{4}$ Bochum Ruhr-University, D-44780 Bochum, Germany\\
% ...
$^{12}$ G.I. Budker Institute of Nuclear Physics SB RAS (BINP), Novosibirsk 630090, Russia\\
% ...
$^{33}$ Johannes Gutenberg University of Mainz, Johann-Joachim-Becher-Weg 45, D-55099 Mainz, Germany\\
$^{34}$ Joint Institute for Nuclear Research, 141980 Dubna, Moscow region, Russia\\
% ...
$^{37}$ Liaoning Normal University, Dalian 116029, People's Republic of China\\
% ...
$^{54}$ Southeast University, Nanjing 211100, People's Republic of China\\
$^{55}$ State Key Laboratory of Particle Detection and Electronics, Beijing 100049, Hefei 230026, People's Republic of China\\
% ...
$^{68}$ University of Science and Technology of China, Hefei 230026, People's Republic of China\\
% ...
$^{71}$ University of Turin and INFN, (A)University of Turin, I-10125, Turin, Italy; (B)University of Eastern Piedmont, I-15121, Alessandria, Italy; (C)INFN, I-10125, Turin, Italy\\
$^{72}$ Uppsala University, Box 516, SE-75120 Uppsala, Sweden\\
% ...
$^{b}$ Also at the Novosibirsk State University, Novosibirsk, 630090, Russia\\
% ...
}
\end{center}
\vspace{0.4cm}
\vspace{0.4cm}
\end{small}
}

\noaffiliation{}

\date{\today}

\begin{abstract}
Based on electron positron collision data collected with the BESIII
detector operating at the BEPCII storage rings, the differential cross
sections of inclusive $\pio$ and $\ks$ production as a function of
hadron momentum, normalized by the total cross section of the $\ee\to$
hadrons process, are measured at six center-of-mass energies from
2.2324 to 3.6710~$\gev$.  Our results with a relative hadron energy
coverage from 0.1 to 0.9 significantly deviate from several
theoretical calculations based on existing fragmentation functions.%,
%especially at lower energies.
\end{abstract}

\maketitle

%%%%%%%%%%%%%%%%%%%%%%%%%%%%%%%%%%%%%%%%%%%%%%%%%%%%%%%%%%%%%%%%%%%%%%%%%%%%%%%%%%
%% introduction
Color confinement at long distance is one of the fundamental
properties of Quantum Chromodynamics (QCD), which is the underlying
theory to describe the strong interactions between quarks and
gluons. Because of confinement, quarks and gluons produced in hard
scattering processes will ultimately become colorless hadrons. The transition
from quarks $q$, antiquarks $\bar{q}$ and gluons $g$ to hadrons $h$,
which occurs at a scale with a small momentum transfer, must be treated
non-perturbatively due to the
large strong-coupling constant $\alpha_s$~\cite{Schmelling:1994py,dEnterria:2022hzv}.
Consequently, the fragmentation function~(FF) $D^h_{q,\bar
  q,g}(z,\mu_F)$ is used to describe the non-perturbative long-distance behavior associated with the hadronization
process~\cite{Metz:2016swz}. Here, $\mu_F$ is the factorization scale used
to factorize the cross section in terms of the convolution of perturbative
hard-part coefficients and non-perturbative fragmentation functions, and
is usually set to be the center-of-mass (c.m.) energy
($\sqrt{s}$) in $\ee$ experiments. The term $z\equiv 2
\sqrt{p_h^2c^2+M^2_hc^4}/\sqrt{s}$ is the relative hadron energy, where
$p_h$ and $M_h$ are the momentum and mass of hadron $h$, respectively.
Single inclusive $\ee$ annihilation, $\ee\to h + X$, where $h$ is an
identified hadron and $X$ represents everything else, provides a
clean way to study FFs~\cite{ParticleDataGroup:2020ssz}.  A typical
experimental observable is
\begin{equation} \frac{1}{\sigma(\ee
    \rightarrow \textmd{hadrons})} \frac{\textmd{d}\sigma(\ee
    \rightarrow h + X)}{\textmd{d}
    p_h},
\label{equation_1}
\end{equation}
where $\sigma(\ee
\rightarrow \textmd{hadrons})$ is the total cross section for $\ee$
annihilation to all possible hadronic final states (referred to as inclusive hadronic events hereafter). At the leading
order in $\alpha_s$, the observable can be interpreted as $\sum_q e_q^2[ D_q^h(z,
  \sqrt{s})+D_{\bar{q}}^h(z,\sqrt{s})]$, where $e_{q}$ is the fractional charge
of the quark $q$.  As summarized in
Ref.~\cite{ParticleDataGroup:2020ssz}, extensive high-precision
measurements of this observable have been made in $\ee$ experiments in
the c.m. energy range above $10~\gev$, but measurements in the region
from 3.6 to 5.2 $\gev$ have statistical uncertainties ranging from
20\% to 50\%~\cite{DASP:1978ftr,PLUTO:1977qwj}.

On the other hand, remarkable progress has been made in the
experimental study of nucleon structure by taking advantage of the
semi-inclusive deep-inelastic scattering (SIDIS) process. Due to its
explicit dependence on both the identified initial and final state
hadrons, the SIDIS process plays a crucial role in probing specific
quark flavors. In particular, the FFs act as the weights
used to perform flavor separation in the initial state. For example,
precise kaon FFs might be the key to
understand the puzzle of the strange-quark polarization inside a
longitudinally polarized proton~\cite{Leader:2011tm,Leader:2014uua}.
The momentum-transfer $Q$ (equivalent to $\sqrt{s}$ in $\ee$
annihilation) for the existing SIDIS data from fixed-target
experiments like the ones at Jefferson Lab, COMPASS and HERMES~\cite{Deur:2018roz} covers the range from
1 to 10 GeV, where only sparse single inclusive $\ee$ annihilation
data exist.  Moreover,
the proposed Electron-Ion Collider (EIC) and the Electron-Ion
Collider in China (EicC)~\cite{Accardi:2012qut,Anderle:2021wcy} would provide very high precision structure function measurements, which, in order to determine precise parton distribution functions (PDFs) for various quark flavors, would put unprecedented requirements on the precision of
FFs for the $Q$ value down to 1 GeV and almost complete $z$ coverage from 0 to 1.
%the proposed Electron-Ion Collider (EIC) and the Electron-ion collider in China (EicC)~\cite{Accardi:2012qut,Anderle:2021wcy} put unprecedented requirements on the precision of FFs for the $Q$ value down to 1 GeV and almost complete $z$ coverage from 0 to 1.
Therefore, comprehensive investigations on the single inclusive annihilation with
identified pion and kaon final states, especially in the region not
well covered by previous data, will provide valuable input for the
nucleon structure study.

In this Letter, the processes $\ee \to \pio/\ks +X$ are studied at six
c.m.~energies from 2.2324 to $3.6710~\gev$, with a $z$ coverage from
0.1 to 0.9.  Data sets used in this research were collected with the
BESIII detector~\cite{BESIII:2009fln} running at BEPCII. The detector
has a geometrical acceptance of 93\% of the 4$\pi$ solid angle for the relatively stable final state particles. It is
based on a superconducting solenoid magnet with a main drift chamber (MDC)
as a central tracking system, plastic scintillators as a
time-of-flight system, CsI crystals as an electromagnetic calorimeter
(EMC) and resistive plate chambers as a muon system.

Monte Carlo (MC) simulations based on {\sc geant4}
software~\cite{GEANT4:2002zbu}, which includes the geometric
description of the BESIII detector and implements the interactions
between the final-state particles and the detector, are used to
optimize the event selection criteria, estimate the number of residual
background events, and determine the correction factors accounting for
the efficiency loss and initial-state radiation (ISR) effects. The
inclusive hadronic events are simulated with the {\sc luarlw}
generator~\cite{luarlwA,luarlwB,BESIII:2021wib}, where among others the signal
processes $\ee\to\pio/\ks+X$ are contained. Background MC samples of
the processes $\ee \to \ee$, $\mu^{+}\mu^{-}$, and $\gamma\gamma$ are
generated by {\sc babayaga3.5}~\cite{CarloniCalame:2000pz}. At
$\sqs=3.6710~\gev$, the $\ee\to\tau^{+}\tau^{-}$ process is simulated
via the {\sc kkmc}~\cite{Jadach:1999vf} generator, in which the decay
of $\tau$ lepton is modeled by {\sc evtgen}~\cite{Lange:2001uf}. The
background events from the two-photon processes are simulated by the
corresponding dedicated MC generators, as detailed in
Ref.~\cite{BESIII:2021wib}. In addition, the beam-associated events
are estimated by a side-band method.

In this analysis, the inclusive hadronic events are selected first,
from which the $\ee \to \pio/\ks +X$ events are identified by
reconstructing a $\pio$ or $\ks$ meson. We begin by removing the dominant background processes, i.e., the $\ee\to\ee$ and $\ee\to\gaga$ events. In each remaining event, the good charged
hadronic tracks are identified with the same selection criteria as described in Ref.~\cite{BESIII:2021wib}, and events with at least two good
tracks are kept for further analysis. For events
with two or three good charged tracks, additional requirements on the charged tracks
and the showers in the EMC are implemented to further suppress the background related to the quantum electrodynamics process. Events with more than three good charged tracks are
retained directly without any additional requirements. The details of
the selection of the inclusive hadronic events are presented in
Ref.~\cite{BESIII:2021wib}.  The numbers of inclusive hadronic
($\Nhadtot$) events and residual background ($\Nbkg$) events, as well
as the integrated luminosity of each data sample, are summarized in
Table~\ref{tab:had}.

\setlength{\tabcolsep}{14pt}
\begin{table}[!htbp]
\setlength{\abovecaptionskip}{5pt}
\setlength{\belowcaptionskip}{10pt}
 \centering
\caption{The integrated luminosities and the total observed hadronic
  and residual background events at various c.m. energy points.}
  \begin{tabular}{c c c c}
  \hline
  \hline
  $\sqrt{s}$~(GeV) & $\mathcal{L}$~($\rm pb^{-1}$) & $\Nhadtot$ & $\Nbkg$\\
  \hline
 2.2324  &   2.645  &  83227  &   2041   \\
 2.4000  &   3.415  &  96627  &   2331   \\
 2.8000  &   3.753  &  83802  &   2075   \\
 3.0500  &   14.89  &  283822 &   7719   \\
 3.4000  &   1.733  &  32202  &   843    \\
 3.6710  &   4.628  &  75253  &   6461  \\
  \hline
  \hline
\end{tabular}
\label{tab:had}
\end{table}

A $\pio$ candidate is reconstructed via the decay $\pio \to \gamma
\gamma$. Photon candidates are identified using showers in the
EMC. The deposited energy of each shower must be greater than
25~$\mev$ in the barrel region ($|\cos\theta| < 0.80$) and 50 $\mev$
in the end-cap region ($0.86 < |\cos\theta| < 0.92$), where $\theta$
is the polar angle defined with respect to the symmetry axis of the MDC.
The shower is required to be separated by more than 10 degrees from the closest
charged track to eliminate those produced by the charged
particles. The difference between the EMC time and the event start
time has to be within $[0, 700]$~ns to suppress electronic noise and
showers unrelated to the event. All combinations of two photons are
used to form $\pio$ candidates. To suppress background due to the
miscombination of photons, requirements are made on the polar angle of
one photon in the helicity frame of the $\pio$ candidate
($\theta_{\gamma}$).  For $\pio$ candidates with momentum less than
$0.3~\gevc$, $|\cos\theta_{\gamma}|$ is required to be less than 0.8,
while for those with momentum larger than $0.3~\gevc$, it must be less
than 0.95. Each $\pi^0$ candidate in an event is counted as a candidate
of the separate inclusive $\pi^0$ event, and the fraction of the observed
hadronic events containing more than one $\pio$ meson varies from 42\% to
50\% at the c.m. energies from $\sqs=2.2324$ to $3.6710~\gev$.
%A test based on pure $J/\psi\to\pipi\pio$
%events demonstrates that those requirements on
%$|\cos\theta_{\gamma}|$
%do not produce peaking contribution to the signal distribution.

A $\ks$ candidate is formed by combining a pair of oppositely charged tracks.
The two tracks are required to satisfy $|\cos\theta|<0.93$. Due to the
relatively long lifetime of the $\ks$ meson, the distance of closest
approach of these charged tracks from the interaction point must be
less than 30 cm along and 10 cm perpendicular to the beam direction.
As a result of the different requirements, the charged track here is not
necessarily one of the good charged tracks identified previously. To increase the
number of events, no particle identification is applied. Further, to
select $\ks$ signal events, the production and decay vertices are
reconstructed, and the decay length between these two vertices is
required to be at least twice its uncertainty. Each $\ks$ candidate
in an event is counted as a candidate of the separate inclusive $\ks$ event. In this measurement, fewer than 1\% observed hadronic events contain more than one $\ks$ meson.

After imposing the above selection criteria, the residual
contributions to the mass spectra of the photon and charged-pion
pairs, $M(\gamma\gamma)$ and $M(\pipi)$, from lepton-pair production,
two-photon processes, and beam-associated events are less than 0.1\%
in both signal processes.  The dominant background is caused by the
miscombinations of the corresponding daughter particles, which are reproduced by the
inclusive hadronic MC samples and well described by a polynomial.

\begin{figure}[!htbp]
\begin{center}
\begin{overpic}[width=0.23\textwidth]{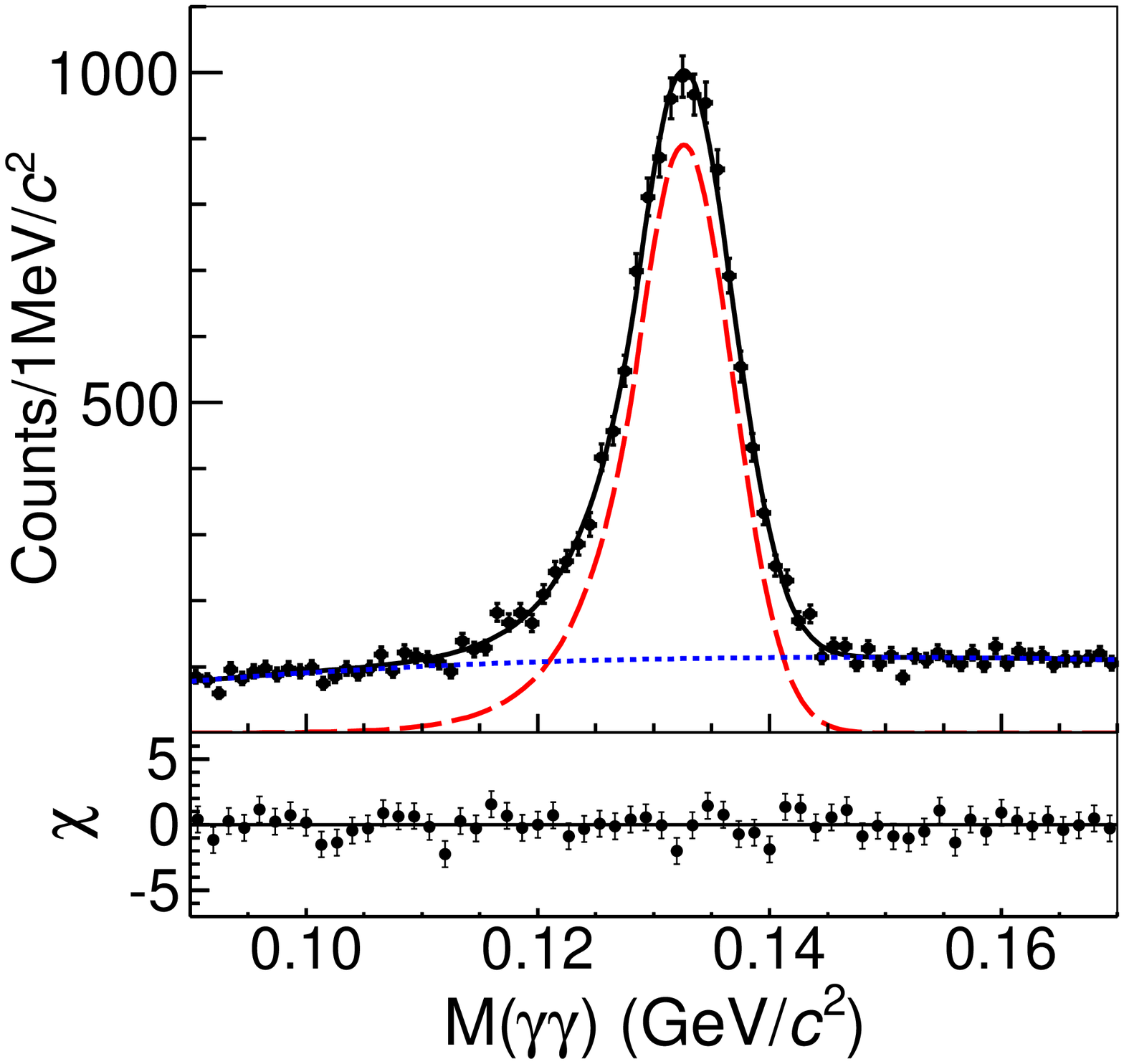}
\end{overpic}
\begin{overpic}[width=0.23\textwidth]{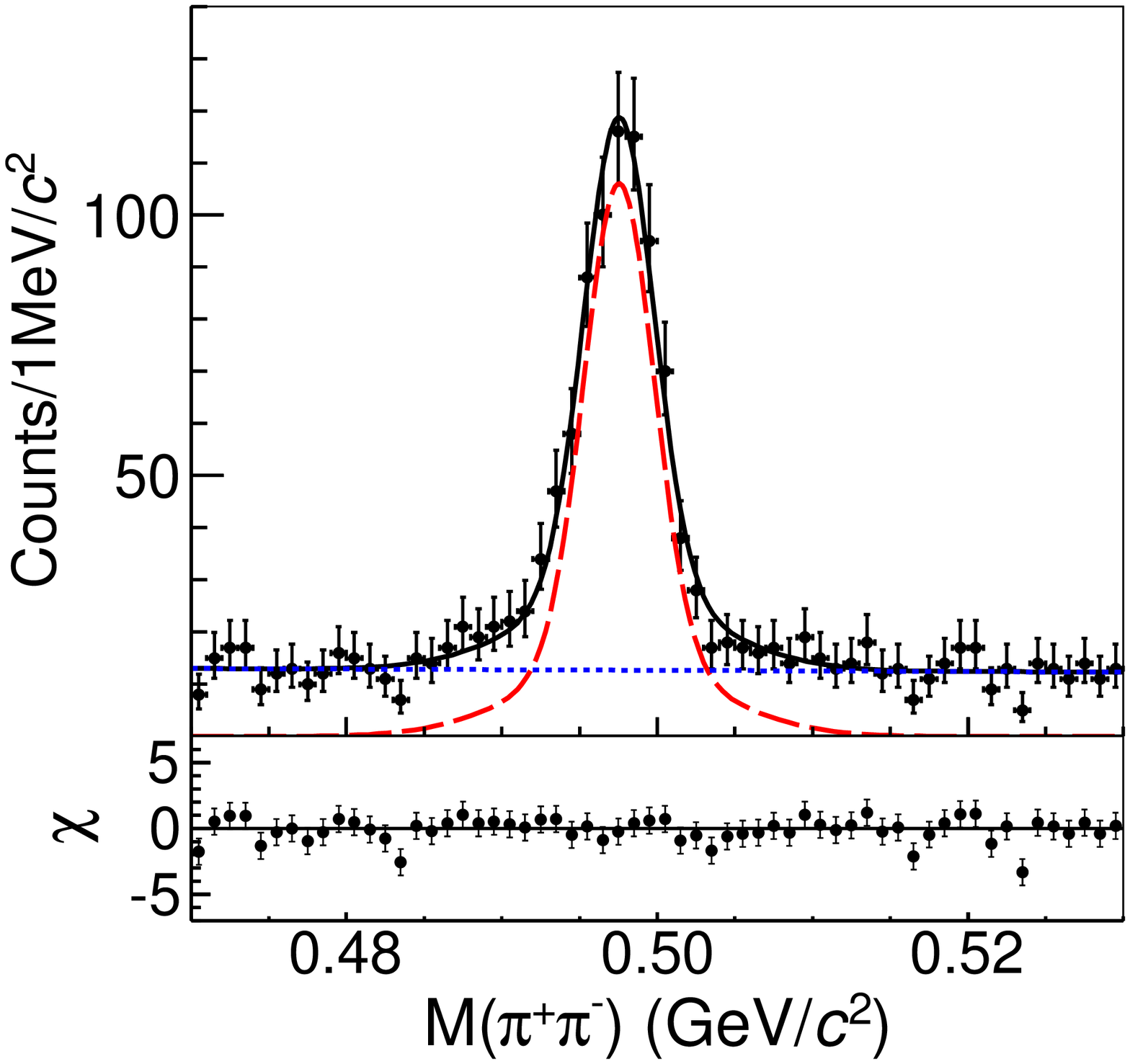}
\end{overpic}
\end{center}
\caption{The M($\gamma\gamma$) and M($\pi\pi$) distributions for $\pio$ (left) and $\ks$ (right) candidates, respectively, with $p_{\gamma\gamma/\pipi}\in(0.4, 0.5)~\gevc$ at $\sqs=2.8000~\gev$. The fit results are overlaid.
The black points with error bars are data. The black solid
curves are the sum of fit functions, while the red dashed and blue
dotted curves represent the signal and background, respectively. The
pull variable $\chi$, defined as the residual between the data and
the total fit function, normalized by the uncertainty of the data,
is shown on the bottom of the figures.}
\label{pioksfit}
\end{figure}

The $M(\gamma\gamma)$ and $M(\pipi)$ spectra are divided into the momentum
intervals with a step of $\Delta p_{\pio/\ks}=0.1~\gevc$, which is 5 times larger than the corresponding momentum resolutions.
Unbinned
maximum likelihood fits are performed on the the
$M(\gamma\gamma)$ and $M(\pipi)$ spectra obtained in each momentum interval to determine the corresponding
numbers of events, i.e., $\Npioobs$ and $\Nksobs$.
For the
$\ee \to \pio+X$ candidates, the signal is described by a Crystal
Ball function~\cite{Skwarnicki:1986xj}, while the background is parameterized by a second-order Chebychev polynomial.  For the $\ee \to \ks +X$ process, the
signal is modeled by a double Gaussian function, and the background is
described by a first-order Chebychev polynomial.
Figure~\ref{pioksfit} illustrates the fit results of the $\pio$ and $\ks$ candidates with
$p_{\gamma\gamma/\pipi}\in(0.4, 0.5)~\gevc$ from the data sample at
$\sqs=2.8000~\gev$. The summary of $\Npioobs$ and $\Nksobs$ obtained in each momentum range at each c.m. energy is given in the Supplemental Material~\cite{suppstuff}.

In practice, the normalized differential cross section for the
inclusive production of any identified hadron, namely
Eq.~(\ref{equation_1}), is determined with
\begin{equation}
\frac{\Nhobs}{\Nhadobs}\frac{1}{\Delta p_h}f_{h},
\label{equation_2}
\end{equation}
where $\Nhadobs=\Nhadtot-\Nbkg$ is the net number of observed
hadronic events in $\ee$ annihilation at a given c.m. energy,
$\Nhobs$ is that of the $\ee \rightarrow h + X$ events within a
specific momentum range $\Delta p_h$, and $f_{h}$ is a correction factor
which accounts for the effects caused by the limited detector
acceptance, the selection criteria, ISR, and vacuum
polarization. Since both $\Nhadobs$ and $\Nhobs$ are obtained from the
same data sample, the integrated luminosity usually used in the cross
section measurement cancels.

In this analysis, the correction factor, $f_h$, is obtained from the
signal MC sample and is given by

\begin{equation}
f_{h}=\frac{\Nbarhtru(\textmd{off})}{\Nbarhadtru(\textmd{off})}\bigg/\frac{\Nbarhobs(\textmd{on})}{\Nbarhadobs(\textmd{on})},
\label{equ:correction_efficiency}
\end{equation}
where the variable ``$\bar{N}$'' denotes the numbers of events
determined from the signal MC sample, either at the detector
observed level, similar to the experimental data, with superscript
``obs'' or at the generation level with superscript ``tru''. The terms
``on'' and ``off'' in the parentheses indicate that the corresponding
quantities are obtained from signal MC samples with and without
simulating the ISR effects, respectively.
$\Nbarhobs(\textmd{on})$ is determined by applying the same fit procedure on the signal MC events as data.
By definition, $f_h$ compensates the event lost caused by the limited acceptance of the BESIII detector in the small polar angle region~\cite{suppstuff}.
In this analysis, the correction factor $f_{h}$ is by far dominated by the detection efficiency.
Calculated results of $f_{h}$ in different momentum ranges for
$\pio$ and $\ks$ are presented in the Supplemental Material~\cite{suppstuff}.

The systematic uncertainties of the normalized differential cross
section measurements mainly originate from the
differences between the signal MC and data samples, the reconstruction
efficiencies of the $\pio$ and $\ks$ candidates, the fits to the events in the
$M(\gamma\gamma)$ and $M(\pipi)$ bins, and the MC simulation model
of the inclusive hadronic events. To estimate the uncertainty caused
by the imperfect simulation of various kinematic variables of the
signal events, all the selection criteria are separately varied to be larger or
smaller than their nominal values by one time their
resolutions, and the maximum changes of the normalized differential
cross sections are taken as the systematic uncertainties.  The
uncertainty in the photon detection efficiency is estimated to be 1\% per
photon~\cite{BESIII:2018rdg}, therefore 2\% is taken as the
systematic uncertainty due to the $\pio$ reconstruction efficiency where the complete correlation between the detection of the two photons is assumed.
The momentum-dependent systematic uncertainties due to the $\ks$ reconstruction efficiency are obtained by applying a dedicated weighting procedure~\cite{BESIII:2015jmz}, that incorporates weights due to the charged particle tracking efficiency and the decay length requirement.
%The systematic uncertainty of the $\ks$ reconstruction efficiency is
%estimated to be 1\%~\cite{BESIII:2015jmz}, which includes those of the
%tracking efficiencies of charged tracks and the decay length requirement.

The uncertainties due to the fits on the events in the $M(\gamma\gamma)$ and
$M(\pipi)$ intervals are examined by using alternative signal and
background shapes. The alternative signal shape of $\ee \to \pio+X$ is
taken as the shape from the signal MC sample, while that for $\ee \to
\ks +X$ is chosen as a single Gaussian function. The alternative
background shapes are obtained by varying the order of the Chebychev
polynomials. The relative differences from the original differential
cross sections are taken as the corresponding systematic
uncertainties.

%Uncertainties associated with the MC mode

The dominant source of systematic uncertainty is the MC simulation
model of the inclusive hadronic events. The generation fractions of
the exclusive processes containing $\pio$ and $\ks$, which make up the
inclusive process, directly affect the correction factors $f_{\pio}$
and $f_{\ks}$. To address the corresponding uncertainty, the {\sc
  hybrid} generator, which was developed in Ref.~\cite{Ping:2016pms}
and improved in Ref.~\cite{BESIII:2021wib}, is used as an alternative
MC model to generate the inclusive hadronic events. In the {\sc
  hybrid} generator, much knowledge of the allowed exclusive processes
in the c.m.~energy region of BESIII has been incorporated, including
measured cross sections and production mechanisms. In addition, a
different simulation scheme of the ISR process is
adopted~\cite{BESIII:2021wib}. The changes of the correction factors
$f_{\pio}$ and $f_{\ks}$ are assigned as the systematic
uncertainties. All these individual systematic uncertainties are
regarded as uncorrelated with each other and are summed in quadrature.

\begin{figure}[!htbp]
\begin{center}
\begin{overpic}[width=0.5\textwidth]{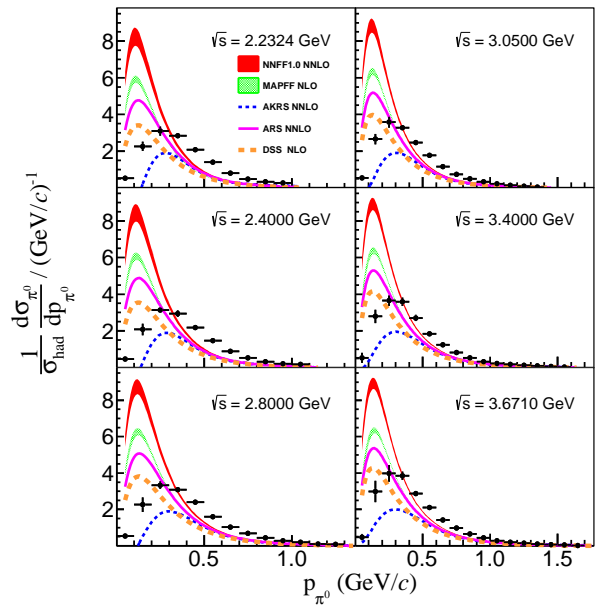}
\end{overpic}
\end{center}
\caption{Normalized differential cross sections of the $\ee \to \pio +
  X$ process. The points with error bars are the measured values, where the
uncertainties are the quadrature sum of the corresponding statistical and systematic uncertainties. The bands or curves in red, green, blue, magenta, and orange denote the NNFF, MAPFF, AKRS, ARS, and DSS calculations, respectively, where only the
former two cover $\pm 1\sigma$ limits. Normalized differential cross sections as function of $z$ are shown in the Supplemental Material~\cite{suppstuff}. }\label{fig:pi0_data_theory}
\end{figure}

\begin{figure}[!htbp]
\begin{center}
\begin{overpic}[width=0.5\textwidth]{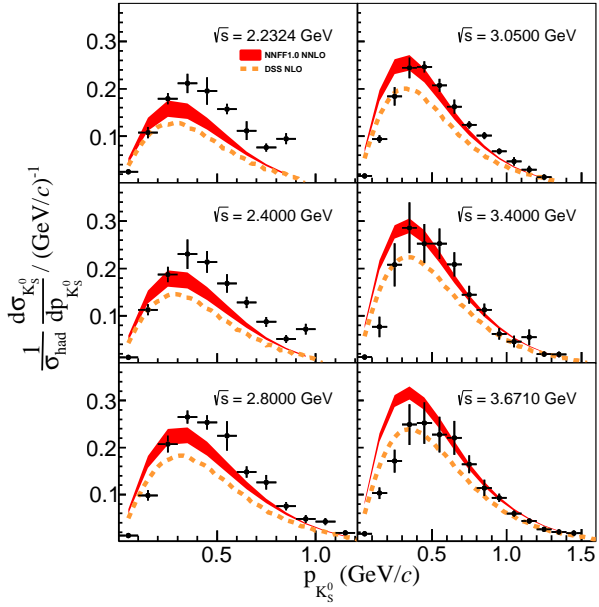}
\end{overpic}
\end{center}
\caption{Normalized differential cross sections of the $\ee \to \ks +
  X$ process. The points with error bars are the measured values, where the
uncertainties are the quadrature sum of the corresponding statistical and systematic uncertainties. The red band shows the theoretical calculation from NNFF with $\pm
1\sigma$ limits and the orange curve denotes the prediction of DDS. Normalized differential cross sections as
function of $z$ are shown in the Supplemental Material~\cite{suppstuff} }\label{fig:KS_data_theory}
\end{figure}

%\textcolor{red}{-----discussions start at here-----}
%%%%%fah
The normalized differential cross sections for the inclusive $\pio$
and $\ks$ production in $\ee$ annihilation at the six c.m.~energies
are tabulated in the Supplemental Material~\cite{suppstuff} and shown
in Fig.~\ref{fig:pi0_data_theory} and Fig.~\ref{fig:KS_data_theory},
respectively.
Figure~\ref{fig:pi0_data_theory} also shows various
theoretical predictions extrapolated from different FFs determined
from existing world
data~\cite{ParticleDataGroup:2020ssz}.
The FFs are obtained with slightly different assumptions and show the
sensitivity of the prediction to assumptions about the behavior at low-$z$
and different $\sqrt{s}$.
% We explicitly choose FFs
%with slightly different assumptions during the extraction in order to
%show the sensitivity of the prediction to such assumptions at low-$z$
%and different $\sqrt{s}$.
ARS~\cite{Anderle:2015lqa}, AKRS~\cite{Anderle:2016czy} and
NNFF1.0~\cite{Bertone:2017tyb} are all obtained from inclusive
annihilation data at NNLO accuracy. However, AKRS includes small-$z$
resummation, NNFF1.0 includes hadron-mass corrections and all of them
have different initial evolution scales and kinematics requirements
on the data. The MAPFF1.0
NLO study~\cite{Khalek:2021gxf} contains low-$Q^2$ data from the lepton-proton fixed-target
experiments at HERMES~\cite{HERMES:2012uyd} and
COMPASS~\cite{COMPASS:2016xvm}.
The DSS NLO calculation~\cite{deFlorian:2007aj} contains low-$Q^2$ data from the lepton-proton fixed-target experiments at HERMES and single inclusive production of proton-proton collisions.
Figure~\ref{fig:KS_data_theory}
shows the comparison of the normalized differential cross section of
the inclusive $\ks$ production with predictions using NNFF1.0 at NNLO precision and DSS at NLO
precision. In these comparisons, the disagreement is observed to depend
on both c.m. energy and hadron momentum. Here, the FFs are further away from
the kinematic region of the original data.   One possible reason for
these discrepancies is that a calculation restricted to the leading twist may not be sufficient at the BESIII energy scale.  It
may also be important to consider quark mass and hadron mass correction
effects~\cite{Accardi:2014qda}, and small-$z$ resummation
effects~\cite{Anderle:2016czy}. Another problem may be with the
extrapolation of the FFs from existing $\ee$ annihilation data at
high energy to the low-energy scale at BESIII. For instance, the predictions using QCD-backward evolution for initial-state PDFs from high to low energies have
been found to deviate from the experimental
measurements~\cite{Caola:2010cy}.  Both
Fig.~\ref{fig:pi0_data_theory} and Fig.~\ref{fig:KS_data_theory} show
that BESIII data can be used to improve the fit procedure to
determine the FFs at the low energy scale. Also, the difference between
primary and secondary processes has to be taken into
account as well. Studies based on the signal MC samples show that the $\rho^{\pm}$ ($K^{*}$) decay has contribution to the inclusive $\pio$ ($\ks$) production in the c.m. energy region of this analysis.
The results
presented in this Letter will be key to explore these possibilites
as well as
help to test
collinear perturbative QCD with data at the
relatively low energy scale.

In summary, we have measured the normalized differential cross
sections of the $\ee \to \pio/\ks + X$ processes, using data samples
collected from $\sqs=2.2324$ to $3.6710~\gev$.  The results obtained
in this work help to fill the region with $\sqs<10~\gev$ where
precision $\ee$ annihilation data have been rarely reported.  The
results provide broad $z$ coverage from 0.1 to 0.9 with precision of
around 3\% at $z\sim 0.4$. These results provide new ingredients for
FF global data fits, in which almost no single inclusive annihilation
data measured in this special energy region has been included.

%\newpage
%\section*{ACKNOWLEDGMENTS}
%%%%%%%%%Summary%%%%%%%%%%%%%%%%%%%%%%%%%%%%%%%%%%%%%%%%%%%%%%%%%%%%%%%%%%%%%%%%%%%%%%%%%%%%%%%%%%%%%%%%%%%%%%%%%%%%%%%%%%
%We would like to thank Hongxi Xing and Daniele Paolo Anderle for very helpful discussions.
The authors would like to thank Daniele Paolo Anderle and Hongxi Xing for their valuable theoretical inputs.
The BESIII Collaboration thanks the staff of BEPCII, the IHEP computing center and the supercomputing center of USTC for their strong support. This work is supported in part by National Key R\&D Program of China under Contracts Nos. 2020YFA0406300, 2020YFA0406400; National Natural Science Foundation of China (NSFC) under Contracts Nos. 11635010, 11735014, 11835012, 11935015, 11935016, 11935018, 11961141012, 12022510, 12025502, 12035009, 12035013, 12192260, 12192261, 12192262, 12192263, 12192264, 12192265, 11335008, 11625523, 12035013, 11705192, 11950410506, 12061131003, 12105276, 12122509, 12205255; the Fundamental Research Funds for the Central Universities, University of Science and
Technology of China under Contract No. WK2030000053; the Chinese Academy of Sciences (CAS) Large-Scale Scientific Facility Program; Joint Large-Scale Scientific Facility Funds of the NSFC and CAS under Contract No. U1832207, U1732263, U1832103, U2032111, U2032105; the CAS Center for Excellence in Particle Physics (CCEPP); 100 Talents Program of CAS; China Postdoctoral Science Foundation under Contracts No. 2019M662152, No. 2020T130636; The Institute of Nuclear and Particle Physics (INPAC) and Shanghai Key Laboratory for Particle Physics and Cosmology; ERC under Contract No. 758462; European Union's Horizon 2020 research and innovation programme under Marie Sklodowska-Curie grant agreement under Contract No. 894790; German Research Foundation DFG under Contracts Nos. 443159800, Collaborative Research Center CRC 1044, GRK 2149; Istituto Nazionale di Fisica Nucleare, Italy; Ministry of Development of Turkey under Contract No. DPT2006K-120470; National Science and Technology fund; National Science Research and Innovation Fund (NSRF) via the Program Management Unit for Human Resources \& Institutional Development, Research and Innovation under Contract No. B16F640076; Olle Engkvist Foundation under Contract No. 200-0605; STFC (United Kingdom); Suranaree University of Technology (SUT), Thailand Science Research and Innovation (TSRI), and National Science Research and Innovation Fund (NSRF) under Contract No. 160355; The Royal Society, UK under Contracts Nos. DH140054, DH160214; The Swedish Research Council; U. S. Department of Energy under Contract No. DE-FG02-05ER41374.

%=============================================================================
%\input{Reference.tex}
%\bibliographystyle{apsrev4-2}
%\bibliography{reference}
%%%%%%%%%%%%%%%%%%%%%%%%%%%%%%%%%%%%%%%%%%%%%%%%%%%%%%%%%%%%%%%%%%%%%%%%%%%%%%

\end{document}